\documentclass[useAMS,referee]{biom}
\providecommand{\newblock}{\hskip .11em\@plus.33em\@minus.07em} % 定义 \newblock
\makeatother

\usepackage{amsmath}
\usepackage{bm}
\usepackage{url}
\usepackage{subcaption}
\usepackage{graphicx}
\usepackage{comment}
\let\bibhang\relax % Clear any prior definition
\usepackage[round,authoryear]{natbib}
\usepackage{amsfonts}
\usepackage{xcolor}
\usepackage{longtable}
\usepackage{multirow}
\usepackage{diagbox}
\usepackage{hyperref}

\def\bfx{{\bmath x}}

\def\bfdelta  {{\nmathbf \delta}}

\def\bfeta    {{\bmath \eta}}

\def\bfdelta  {{\bmath \delta}}

\def\bfbeta   {{\bmath\beta}}

\title[Bayesian Extrapolation Design Exposure-Response Curve Comparison]{A Novel Bayesian Extrapolation Design for Assessing Equivalence in Exposure-Response Curves between Pediatric and Adult Populations
}

\author
{Zhongheng Cai \\
Department of Biostatistics, St. Jude Children's Research Hospital, Memphis, USA
\and
Lian Ma \\
Createrna Science and Technology, Gaithersburg, Maryland, USA
\and
Jingjing Ye\email{jingjing.ye@beigene.com} \\
Global Statistics and Data Science at BeiGene, Washington DC, USA
\and
Haitao Pan\email{Haitao.Pan@stjude.org}\\
Department of Biostatistics, St. Jude Children's Research Hospital, Memphis, USA
}

\begin{document}

%  This will produce the submission and review information that appears
%  right after the reference section.  Of course, it will be unknown when
%  you submit your paper, so you can either leave this out or put in 
%  sample dates (these will have no effect on the fate of your paper in the
%  review process!)

%\date{{\it Received October} 2007. {\it Revised February} 2008.  {\it
%Accepted March} 2008.}

%  These options will count the number of pages and provide volume
%  and date information in the upper left hand corner of the top of the 
%  first page as in published papers.  The \pagerange command will only
%  work if you place the command \label{firstpage} near the beginning
%  of the document and \label{lastpage} at the end of the document, as we
%  have done in this template.

%  Again, putting a volume number and date is for your own amusement and
%  has no bearing on what actually happens to your paper!  

%\volume{64}
%\pubyear{2008}
%\artmonth{December}

%  The \doi command is where the DOI for your paper would be placed should it
%  be published.  Again, if you make one up and stick it here, it means 
%  nothing!

%\doi{10.1111/j.1541-0420.2005.00454.x}

%  This label and the label ``lastpage'' are used by the \pagerange
%  command above to give the page range for the article.  You may have 
%  to process the document twice to get this to match up with what you 
%  expect.  When using the referee option, this will not count the pages
%  with tables and figures.  

%\label{firstpage}

%  put the summary for your paper here

\begin{abstract}
Development of effective treatments in pediatric population poses unique scientific and ethical challenges in addition to the small population. In this regard, both the U.S. and E.U. regulations suggest a complementary strategy, pediatric extrapolation, based on assessing the relevance of existing information in the adult population to the pediatric population. The pediatric extrapolation approach often relies on data extrapolation from adults, contingent upon evidence of similar disease progression, pharmacology and clinical response to treatment between adult and children. Similarity evaluation in pharmacology is usually characterized through the exposure-response relationship. Current methodologies for comparing exposure-response (E-R) curves between these groups are inadequate, typically focusing on isolated data points rather than the entire curve spectrum \citep{Zhang:2021}. To overcome this limitation, we introduce an innovative Bayesian approach for a comprehensive evaluation of E-R curve similarities between adult and pediatric populations. This method encompasses the entire curve, employing logistic regression for binary endpoints. We have developed an algorithm to determine sample size and key design parameters, such as the Bayesian posterior probability threshold, and utilize the maximum curve distance as a measure of similarity. Integrating Bayesian and frequentist principles, our approach involves developing a method to simulate datasets under both null and alternative hypotheses, allowing for type I error and type II error control. Simulation studies and sensitivity analyses demonstrate that our method maintains a stable performance with type I error and type II error control.
\end{abstract}

%  Please place your key words in alphabetical order, separated
%  by semicolons, with the first letter of the first word capitalized,
%  and a period at the end of the list.
%


%  As usual, the \maketitle command creates the title and author/affiliations
%  display 

\maketitle

%  If you are using the referee option, a new page, numbered page 1, will
%  start after the summary and keywords.  The page numbers thus count the
%  number of pages of your manuscript in the preferred submission style.
%  Remember, ``Normally, regular papers exceeding 25 pages and Reader Reaction 
%  papers exceeding 12 pages in (the preferred style) will be returned to 
%  the authors without review. The page limit includes acknowledgements, 
%  references, and appendices, but not tables and figures. The page count does 
%  not include the title page and abstract. A maximum of six (6) tables or 
%  figures combined is often required.''

%  You may now place the substance of your manuscript here.  Please use
%  the \section, \subsection, etc commands as described in the user guide.
%  Please use \label and \ref commands to cross-reference sections, equations,
%  tables, figures, etc.
%
%  Please DO NOT attempt to reformat the style of equation numbering!
%  For that matter, please do not attempt to redefine anything!
	\section{Introduction}
Pediatric drug development presents unique challenges due to limited patient populations, ethical constraints on clinical trials, and the need for age-appropriate dosing strategies. Conducting large-scale clinical studies in children is often infeasible, making it difficult to establish robust efficacy and safety profiles.  Regulatory agencies, including the U.S. Food and Drug Administration (FDA) and the European Medicines Agency (EMA), advocate for pediatric extrapolation, a methodology that leverages existing adult data to inform pediatric dose selection and efficacy assessment. This approach accelerates drug development while minimizing unnecessary pediatric trials, ensuring that children receive timely access to safe and effective medications
	
	The International Council for Harmonisation (ICH) guidelines \citep{ICHE11A:2022} provide a regulatory framework for pediatric extrapolation, emphasizing the importance of assessing similarities in disease progression, drug pharmacology, and treatment response between adults and children. Despite these advancements, no standardized statistical methodology exists for quantitatively evaluating exposure–response (E–R) similarity, an essential factor in pediatric dose selection and extrapolation decisions. Current regulatory recommendations encourage systematic extrapolation strategies; however, there remains no clear methodological consensus on how to define and assess E–R similarity in a rigorous and reproducible manner. For a recent review of pediatric extrapolation in U.S. drug development, see \cite{Ye:2023}. 

 Traditionally, pediatric dose finding has relied on E–R relationships to establish appropriate dosing regimens. While historical approaches based on visual inspection of E–R curves and pointwise statistical comparisons provide qualitative insights, they introduce subjectivity and may fail to capture the full spectrum of differences between adult and pediatric E-R profiles, particularly across the entire dose-response continuum. \cite{Zhang:2021} advanced this field by proposing a noninferiority-based framework for assessing E–R similarity, though their method is limited to selecting specific exposure levels rather than evaluating similarity across the entire curve.

 Recent methodological work has explored more sophisticated approaches. Pharmacokinetic/pharmacodynamic (PK/PD) modeling provides a mechanistic framework for predicting pediatric responses based on adult data, while frequentist methods using maximum deviation metrics offer quantitative tools for comparing regression models \citep{Dette:2018,Dette:2024}. However, these approaches do not constitute a formal Bayesian design that enables efficient borrowing of adult data while controlling the type I error.

To address these gaps, we propose the Pediatric Bayesian Extrapolation Design (PED), a comprehensive statistical framework for evaluating whole-curve E–R similarity between pediatric and adult populations. PED integrates prior information from adult trials, reducing the required pediatric sample size while ensuring type I error control and statistical power. Our approach provides a structured Bayesian framework for determining whether the pediatric E–R curve sufficiently resembles the adult curve, enhancing the rigor and efficiency of extrapolation decisions. Our work introduces several key innovations:
\begin{enumerate}
    \item[1.] Development of the Robust Elicited Points Prior (REPP), a novel Bayesian prior specifically tailored for pediatric extrapolation that flexibly incorporates adult E–R data information.
    \item[2.] Implementation of a whole-curve similarity assessment based on the maximum deviation metric, providing a global measure of similarity between pediatric and adult E–R curves.
    \item[3.] Introduction of a Bayesian decision framework that integrates type I and type II error control in pediatric extrapolation settings.
    \item[4.] Introduction of a measure that selects performance-stable design parameters, ensuring reproducibility and efficiency in pediatric trial planning.
\end{enumerate}

The remainder of this manuscript is structured as follows. Section 2 introduces the Bayesian extrapolation framework, detailing the derivation of REPP and the generation of the pediatric coefficients under null and alternative hypotheses. Section 3 presents an algorithm for study design optimization, including type I error control and statistical power considerations. Section 4 illustrates the application of PED through a real-world case study in a pediatric clinical setting. Section 5 conducts sensitivity analyses, assessing the robustness of PED across various parameter settings. Section 6 discusses broader implications for pediatric dose finding, regulatory applications, and future research directions. By extending Bayesian inference to whole-curve pediatric extrapolation, our methodology enhances statistical rigor, efficiency, and reliability, offering a structured Bayesian design of extrapolation for drug developers and regulatory decision-makers. This approach has the potential to streamline pediatric drug development, optimize dose selection, and improve the efficiency of pediatric trial designs, ultimately facilitating faster regulatory approval and access to therapies for children.

	\section{Pediatric Bayesian Extrapolation Design (PED)}
	To assess the similarity of two E-R curves, we employ the equivalence hypothesis testing framework proposed by \cite{Dette:2018}. The objective is to determine whether the difference between two curves can be considered "similar." Traditionally, in the field of extrapolation, this involves selecting specific exposure levels/range and comparing responses at these levels, either visually or through selected summary statistics, comparing mean or median responses of such as AUC, Cmax, Cmin, or Caverage between adult and pediatric groups. However, these methods may not be adequate. For instance, while establishing similarity between the overall E-R curves implies similarity in the selected exposure levels or summary statistics, the reverse does not necessarily hold true. Our goal is to assess the "similarity" of the entire curves in the interval of interest, which necessitates a clear definition of what constitutes similarity. Following \cite{Dette:2018,Dette:2024}, a maximal deviation between the two curves is considered as a measure of similarity, for which corresponding results are substantially harder to derive (e.g., if the maximal part between two curves can be tested for similarity, the rest part of the curves should also be similar) and meanwhile the methodology based on this measure is demonstrated to be a more powerful than the other tests \citep{Dette:2018}. Thus, the one-sided hypotheses are defined as follows:
	\begin{equation}
		H_0:\max_{x\in[a,b]}\left\{f_{adult}(x,\bfbeta_{adult})-f_{ped}(x,\bfbeta_{ped})\right\}\geq\epsilon_H,\\
		H_1: \max_{x\in[a,b]}\left\{f_{adult}(x,\bfbeta_{adult})-f_{ped}(x,\bfbeta_{ped})\right\}<\epsilon_H.
		\label{hypo}
	\end{equation} 
	where $f_{ped}(x, \bfbeta_{ped})$ denotes the probability curve for the binary endpoint in the pediatric population, and $f_{adult}(x, \bfbeta_{adult})$ denotes the fitted probability curve in the adult population. The parameter $\epsilon_H$ denotes the threshold used to assess the similarity between these curves. In contrast to the hypothesis testing approach described in \cite{Dette:2018}, a one-sided test is employed in this study, reflecting the practitioner's primary concern with cases where the pediatric response is lower than the adult response. Similarity is considered to hold if the maximum difference between the adult and pediatric responses, defined as $\max_{x\in[a,b]}\left\{ f_{adult}(x, \bfbeta_{adult}) - f_{ped}(x, \bfbeta_{ped}) \right\}$, is less than $\epsilon_H$. The whole exposure range, defined as [A, B], is determined by adult studies and represents the full range of observed exposures or doses. However, in our hypothesis, we focus on a subset interval $[a,b]\subseteq[A,B]$. This interval [a, b] typically corresponds to the region where the exposure-response relationship exhibits a steeper gradient, indicating a more pronounced therapeutic effect in adult population. The selection of [a, b] is primarily driven by clinical considerations and is determined by the clinical team based on their expertise and understanding of the therapeutic window. 

 We use the logistic curve to model the ER curve, 

\begin{equation}
		f_{\text{ped}}(x, \boldsymbol{\beta}_{\text{ped}}) = \frac{\exp(\beta_{\text{ped},1} + \beta_{\text{ped},2} x)}{1 + \exp(\beta_{\text{ped},1} + \beta_{\text{ped},2} x)}; \quad f_{\text{adult}}(x, \boldsymbol{\beta}_{\text{adult}}) = \frac{\exp(\beta_{\text{adult},1} + \beta_{\text{adult},2} x)}{1 + \exp(\beta_{\text{adult},1} + \beta_{\text{adult},2} x)},
        \label{eq_gene}
	\end{equation}
Here, $\bfbeta^{'}_{\text{ped}}=(\beta_{\text{ped,1}},\beta_{\text{ped,2}})$ and $\bfbeta^{'}_{\text{adult}}=(\beta_{\text{adult,1}},\beta_{\text{adult,2}})$ and $x\in [A,B]$. The logistic curve has been extensively employed in medical research to model response probabilities. For instance, in the case of infliximab, it characterizes the relationship between drug concentrations and the primary clinical response endpoint. The FDA approved infliximab for pediatric use based on an extrapolation design \citep{Mulberg:2022}. Similarly, logistic regression analysis has been applied to pooled data from both pediatric and adult patients for darunavir. The FDA review confirmed comparable relationships between drug exposure—measured by the $\log_{10}$ IQ ratio or darunavir trough concentration ($C_{0h}$)—and response, defined as the proportion of subjects achieving virologic success, across these populations \citep{FDA:2008}.
 
	We adopt the following Bayesian decision rule to accept $H_1$, thereby claiming that the two E-R curves are similar enough to allow extrapolation between the two groups:
	\begin{equation}
		P(\text{ reject }H_0|Data)>\epsilon_{bayes}
		\label{decision}
	\end{equation}
	where $\epsilon_{\text{bayes}}$ is \textit{a tuning parameter} to represent the probability threshold to declare similarity. 

 To the best of our knowledge, no previous studies have employed Bayesian methods for hypothesis testing based on the maximal deviation between two curves in the context of pediatric drug development. We propose a novel approach that extends the work by \cite{Dette:2018} to a Bayesian hypothesis testing framework and introduces a method for sample size determination that considers both type I error rate and power.
 
	A key innovation in our method is twofold. First, we develop a new type of prior distribution for the coefficients of logistic curve, Robust Elicited Points Prior (REPP), in our model. The REPP is formulated as a mixture that combines non-informative priors with informative priors which may derive from the adult data information. Second, we propose an approach to generate data under both the null and alternative hypotheses. This method is specifically tailored to the problem of how to generate coefficients under the maximal deviation metric due to its infinite nature when $H_0$ or $H_1$ is true. It hinders us to generate the dataset under either $H_0$ or $H_1$ since different value of  $\bfbeta_{\text{ped}}$ will result in different operating characteristics and there is no available guidance to choose $\bfbeta_{\text{ped}}$. To tackle this problem, we propose a novel method that categorizes coefficients and employs a sampling strategy under two hypotheses, which is essential for controlling type I error and type II error. The main idea is that we find a quantity $\eta$, which is monotone with (3), that can divide the whole space of the $\bfbeta_{\text{ped}}$ satisfying (1). Then, we can assign the pre-selected weight, which is carefully chosen to balance the control of type I error and type II error in worst-case scenarios. The details are provided in the following subsections.
	\subsection{Robust Elicited Points Prior (REPP)}
Various information-borrowing priors have been developed in the literature, including the power prior \citep{Chen:Ibra:2000}, commensurate prior \citep{Hobbs:2011}, and robust meta-analytic prior \citep{Schmidli:2011}. However, to the best of our knowledge, all the proposed priors are specifically directly on the coefficients of the response curve. In this study, we introduce the Robust Elicited Points Prior (REPP), a novel approach that derives priors for curve coefficients based on exposure information of selected points within the exposure interval of interest. This method integrates expert knowledge without directly imposing fixed priors on the coefficients themselves. The REPP is constructed through three main steps:
\begin{enumerate}
    \item[1.] \textbf{Selection of Exposure Points:} Select a set of exposure points within the interval $[a,b]$, our interested sub-interval, where experts can infer the information of the deviation, which is defined as the difference of log odds ratio between the pediatric and the fitted adult curve from the completed adult trial(s). The information could be the mean and variance or the 25th percentile, median, and 75th percentile of the deviation.
     \item[2.] \textbf{Generation of Synthetic Data:} Generate synthetic data at the chosen points and obtain posterior samples of the coefficients for the pediatric curve. We use these samples to form the informative component of the REPP.
      \item[3.] \textbf{Combination with Non-Informative Priors:} Combine the informative component with a non-informative normal distribution with a large variance.
\end{enumerate}
Specifically, in Step 1, exposure points within the interval [a,b] are selected to facilitate the elicitation of pediatric response information. Considering the logistic curve with two unknown parameters, we recommend selecting three points, $\bfx=(x_1,x_2,x_3)$, where experts can readily infer statistical information about the deviation. We consider the following two scenarios:
\begin{enumerate}
    \item[(i)] The expert can provide the mean and variance of the deviation at the selected points;
    \item[(ii)] The expert can provide the quantile information of the deviation at the selected points.
\end{enumerate}

This information in two scenarios may be obtained by the experts from clinical pharmacological data in similar drugs. In Scenario (i), if experts provide the mean and variance at these selected points, for pediatrics, a simple assumption of the distribution for the deviation follows a normal distribution. In Scenario (ii), if quantile information is provided, determining the corresponding normal distribution is less straightforward, and we may employ the Sheffield Elicitation Framework (SHELF)\citep{Oakley:2023}. SHELF provides protocols, templates, and guidance for expert elicitation, utilizing methods such as quantile judgments (e.g., median, tertiles) or probability judgments (e.g., probability of an uncertain quantity within a specified interval). In the current context, we may ask experts: 

\textit{“Based on your knowledge, what are the 25th percentile, median, and 75th percentile of the deviation of log odds ratio between the pediatric and fitted adult curve at the selected points?” }

Once these quantiles are obtained, normal distributions can be fitted using a least squares approach, adjusting parameters to align the quantiles of the fitted normal distribution as closely as possible with expert judgments \citep{Oakley:2023}.

For Step 2, we utilize the mean-shift model to describe the relationship between the pediatric exposure-response (E-R) curve and the adult E-R curve:
\begin{equation}
		\bfx^{'}\bfbeta_{ped}=\bfx^{'}\bfbeta_{adult}+devation(x),x\in [a,b]
		\label{gene:1}
	\end{equation}
where $devation(x)\sim \mathcal{N}(B(x),\sigma^2_B(x))$, $B(x)$ and $\sigma_B^2$ are the known mean and variance, respectively. In our case, we focus on three points, $\bfx=(x_1,x_2,x_3)$. From equation \eqref{gene:1}, we infer that pediatric information can be regarded as a fluctuation around the adult information. This assumption is reasonable because, in developing an extrapolation plan, there should already be scientific evidence suggesting similarity between the two groups in terms of efficacy, as reflected by the dose-response curves. In the Step (1),  we focus on three points, $\bfx=(x_1,x_2,x_3)$, to generate the dataset. We follow these steps to obtain the informative part of Robust Elicited Points Prior (REPP) for
$\bfbeta_{\text{ped}}$
 \begin{enumerate}
    \item[(1)] Generate synthetic data points at each $\bfx=(x_1,x_2,x_3)$, with the distributions of $devation(x_i)$ for $i = 1, 2, 3$, we can generate the data points for each $x_i$ and transfer $\bfx^{'}\bfbeta_{ped}$ into probability using \eqref{eq_gene}.
     \item[(2)] Solve the minimization problem defined in \eqref{eq:2} using a Bayesian method to obtain the posterior sample of each component of $\bfbeta_{\text{ped}}$.
      \item[(3)] Approximate the posterior sample using a mixture of two normal distributions, which forms the informative part of the REPP.
\end{enumerate}
We generate a synthetic data with 1000 data points, $\{(x_i,y_{ij})|i=1,2,3;j=1,\cdots,1000\},$ for each chosen point $x_i$ in $\bfx=(x_1,x_2,x_3)$. We estimate $\bfbeta_{\text{ped}}$ based on the following minimization problem:
\begin{equation}
		\min_{\bfbeta_{\text{ped}}} \sum_{i=1}^{3} \sum_{j=1}^{1000} \left(y_{ij} -f_{ped}(x_i,\bfbeta_{\text{ped}})\right)^2
		\label{eq:2}
	\end{equation}
where $\bfbeta_{\text{ped}}=(\beta_{\text{ped,1}},\beta_{\text{ped,2}})$. We may obtain the estimated $\bfbeta_{\text{ped}}$ by solving \eqref{eq:2}. However, what we need is the prior distribution of $\bfbeta_{\text{ped}}$ instead of its estimator. Hence, we may resort to Bayesian method to solve \eqref{eq:2}. We adopt the following assumptions:
\begin{align}
		y_{ij} &\sim \mathcal{N}\left(f_{ped}(x_i,\bfbeta_{\text{ped}}), \sigma_F^2\right), \quad \text{for } i = 1, 2, 3 \text{ and } j = 1, \ldots, 1000, \nonumber \\
		\beta_{\text{ped}, k} &\sim \mathcal{U}(-\infty, +\infty), k=1,2\nonumber
	\end{align}
	where $\mathcal{U}$ denote the uniform distributions, respectively. With likelihood of $y_{ij}$ and the prior of $\beta_{\text{ped,k}}$, we may find that the kernel of the posterior of $\bfbeta_{\text{ped}}$ is \eqref{eq:2}. Hence, the posterior mode of $\bfbeta_{\text{ped}}$ serves as the solution to \eqref{eq:2}. In terms of fully Bayesian analysis, for the nuisance parameter, $\sigma_F^2$, a Jeffreys prior is applied. Note that the likelihood and prior of $\beta_{\text{ped},k}$ are utilized solely to derive the posterior sample of $\beta_{\text{ped},k}$. This approach is akin to the use of "functional" priors in Bayesian LASSO \citep{Park:Casella:2008} and Bayesian quantile analysis \citep{Yu:Moye:2001}. We derive the informative part of the prior distribution for the coefficients $\bfbeta_{\text{ped}}$ using a mixture of normal distributions and the number of mixture components are two. For the non-informative component, we use $N(0,100^2)$.   Letting $w$ denote the weight of the informative part, the prior distribution for $\bfbeta_{\text{ped}}$ is given by:
\begin{equation}
		\pi^f(\beta_{\text{ped}, k}) = (1-w) \mathcal{N}(0, 100^2) + w \left[ p \mathcal{N}(\mu_{1k}, \sigma^2_{1k}) + (1 - p) \mathcal{N}(\mu_{2k}, \sigma^2_{2k}) \right]
		\label{prior}
	\end{equation}
where $p$ denotes the weight of the mixture normal distribution, and $\mu_{1k}, \mu_{2k}, \sigma^2_{1k}$, and $\sigma^2_{2k}$ represent the means and variances of the two normal distributions derived from the posterior sample. All parameters can be estimated using the \texttt{flemix} package in R. $w$ measures the information we borrow from the adult population. 

 By employing these distributions derived from adult data and expert knowledge, we can construct priors for the coefficients of the pediatric exposure-response curve. This method bridges the gap between fully data-driven approaches and purely elicitation-based methods for curves comparison, offering a robust solution for information borrowing in pediatric clinical trial design.
	\subsection{Data Generation}
In this section, we propose a framework for generating simulated data under the two hypotheses outlined in \eqref{hypo}. We consider the coefficient vector $\bfbeta_{\text{ped}}$ satisfying 
\begin{equation}
		\max_{x\in[a,b]}\left\{f_{\text{adult}}(x, \boldsymbol{\beta}_{\text{adult}}) - f_{\text{ped}}(x, \boldsymbol{\beta}_{\text{ped}})\right\}= \delta
		\label{eq_2.2}
	\end{equation}
	In this context, under $H_1$, the value of $\delta$ represents a margin within the range $[0, \epsilon_H)$, which quantifies the similarity between the two exposure-response (E-R) curves. Conversely, under the null hypothesis $H_0$ is true, $\delta$ is set to $\epsilon_H$, marking the critical threshold for hypothesis testing. It is important to note that, for a fixed $\delta$, there are infinitely many $\bfbeta_{\text{ped}}$ that satisfy equation \eqref{eq_2.2}. The main challenge is determining an approach to sample $\bfbeta_{\text{ped}}$  satisfying this constraint. To address this, we focus on $\bfbeta_{\text{ped}}$  such that $f_{\text{ped}}(x,\bfbeta_{\text{ped}})\leq f_{\text{adult}}(x,\bfbeta_{\text{adult}})$ for $x\in[a, b]$.  This restriction is justified by clinical pharmacology, where the focus is on intervals where the adult curve is greater than or equal to the pediatric curve.

\begin{figure}
    \centering
    \begin{subfigure}[b]{0.48\textwidth}
        \centering
        \includegraphics[width=\textwidth]{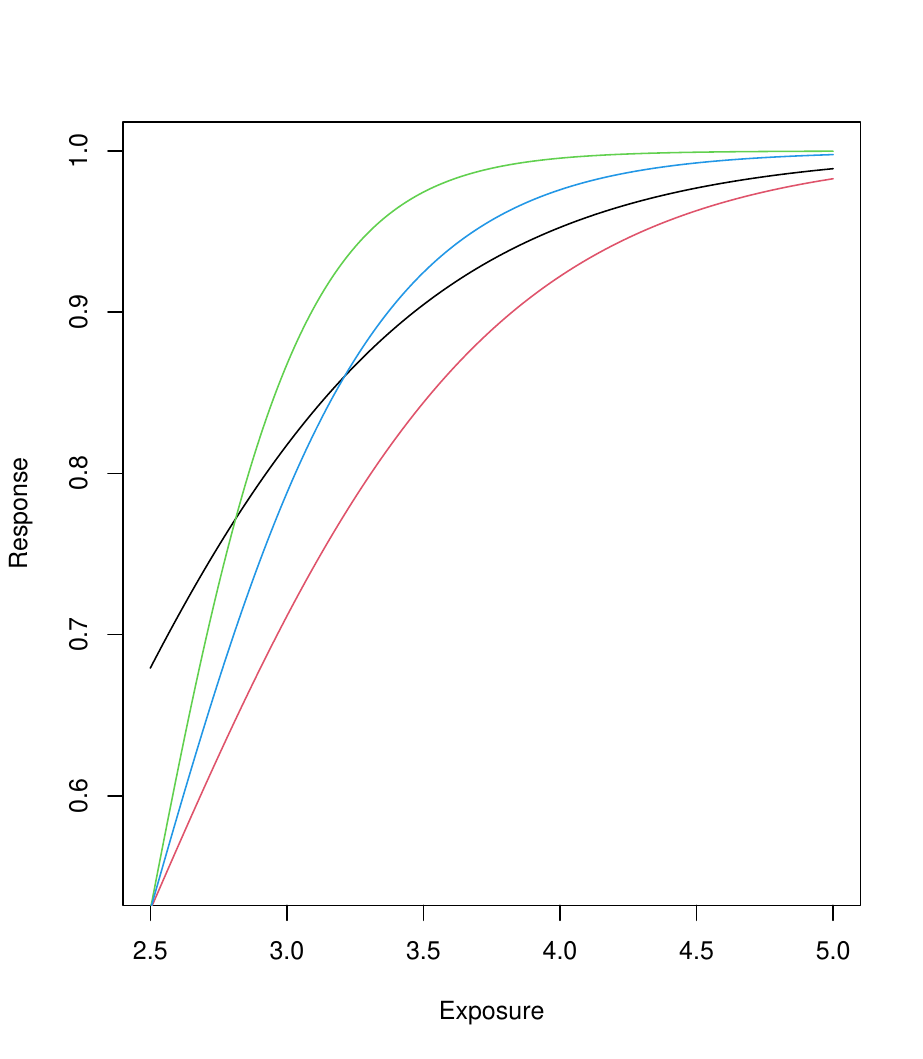}
        \caption{Black line: adult E-R curve; The lines with green, blue and red colors are the pediatric E-R curves with same maximal distance.}
        \label{fig:1}
    \end{subfigure}
    \hfill
    \begin{subfigure}[b]{0.48\textwidth}
        \centering
        \includegraphics[width=\textwidth]{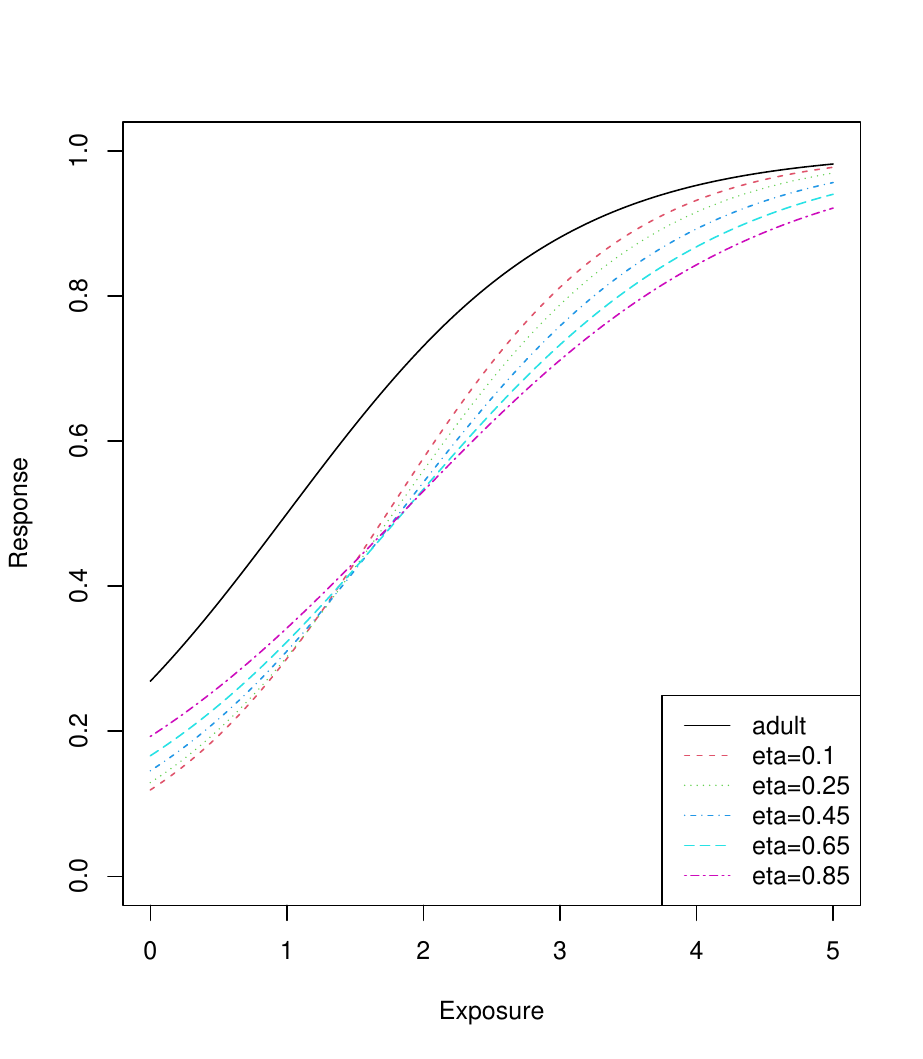}
        \caption{Black line: adult E-R curve. Colored and dashed lines are the pediatric E-R curves with different $\eta$.}
        \label{fig:3}
    \end{subfigure}
    \caption{Comparison of pediatric E-R curves under different settings.}
\end{figure}

To illustrate this concept, we provide an example by setting $\epsilon_H=0.2$. Figure \ref{fig:1} displays scenarios where  $f_{\text{ped}}(x,\bfbeta_{\text{ped}})\leq f_{\text{adult}}(x,\bfbeta_{\text{adult}})$ holds in  different subintervals. The black line represents the adult E-R curve, while the three colored lines represent fitted pediatric curves, each maintaining a maximal distance of 0.15 when $x=2.5$. Although all three curves satisfy the criterion that the maximal distance is less than $\epsilon_H$, indicating potential similarity, the red curve requires further evaluation as it consistently lies below the adult curve. In other words, if our interested exposure range is from 3.5 and beyond, the green and blue lines show that efficacy of pediatrics is not just similar and even better than the adult, but the red line is unconclusive in terms of similarity, though these three lines have the same maximal distance. That is the reason why we will focus on the red line or equivalent the constraints of $f_{\text{ped}}(x,\bfbeta_{\text{ped}})\leq f_{\text{adult}}(x,\bfbeta_{\text{adult}})$  for $x\in[a, b]$ to generate the data set.  To be mentioned, despite this constraint, the number of $\bfbeta_{\text{ped}}$ remains infinite.

In view of this case, there are still two main challenges arising in sampling a reasonable $\bfbeta_{\text{ped}}$: (i), there is no practical method to store all possible $\bfbeta_{\text{ped}}$ satisfying (7) due to its infinity, and (ii), under the null hypothesis ($H_0$), specific $\bfbeta_{\text{ped}}$ values may yield a high Type I error rate, whereas under the alternative hypothesis ($H_1$), certain values may compromise statistical power. Therefore, it is imperative to assign greater weight to $\bfbeta_{\text{ped}}$ values that represent the most adverse scenarios. The sampling framework for generating $\bfbeta_{\text{ped}}$ values adheres to the following principle: under $H_0$, priority is given to values with elevated Type I error, while under $H_1$, emphasis is placed on values associated with elevated Type II error.

Suppose that we have one quantity $\eta\in[0,1]$ with one-to-one relationships with respect to $\bfbeta_{\text{ped}}$ given $\delta$ and it has a decreasing trend with $P(\text{ reject }H_0|Data)$. That quantity will help us divide the space of $\bfbeta_{\text{ped}}$ efficiently, since it is possible to discretize the space of $\eta$, such as $0.1,\cdots,0.9$, and store the corresponding $\bfbeta_{\text{ped}}$. And our weight-imposing method on $\eta$ is as follows: when $H_0$ holds, we impose a decreasing weight on $\eta$; when $H_1$ holds, we impose an increasing weight on $\eta$. Specifically, a decreasing weight on $\eta$ means that a smaller $\eta$ (e.g, 0.1, 0.2, $\cdots$) will be assigned a larger probability than a larger $\eta$ (e.g., 0.9, 0.8,$\cdots$); therefore, under $H_0$, this specific weight-imposing approach will prone to have a larger $P(\text{ reject }H_0|Data)$, e.g, a larger type I error, which is conservative and follows the above guiding principle. An increasing weight can be interpreted in a same way.

Based on previous discussion, given $\delta$ in \eqref{eq_2.2}, the weight imposing method will help us to prepare for the worst cases under $H_0$ and $H_1$. Under $H_0$, we have $\delta=\epsilon_H$, hence, the procedure of sampling $\bfbeta_{\text{ped}}$ has already described as above. When $H_1$ holds, and if we  want to generate $\bfbeta_{\text{ped}}$, one issue should be solved is that currently $\delta\in[0,\epsilon_H)$, that is, $\delta$ has infinite values to choose different from when $H_0$ holding, Our algorithm will integrate the $\delta$ (details see later), so we have to assign a distribution to $\delta$. One option is uniform distribution at $[0,\epsilon_H]$, however, this regards all possible values as equal and we don’t think the values are close to 0 or $\epsilon_H$  are reasonable since it indicates that the two E-R is similar or dissimilar to a great extent , based on this consideration, we opt to have a distribution centered with its mode around $\epsilon_H/2$, for example, the mode could be 0.3$\epsilon_H$, 0.5$\epsilon_H$ or 0.7$\epsilon_H$.

Our next objective is to derive the formula for $\eta$. Let $S_{D(x;\boldsymbol{\beta}_{\text{ped}})}$ denote the area of $D(x;\boldsymbol{\beta}_{\text{ped}})$, where  
\[
D(x;\boldsymbol{\beta}_{\text{ped}}) = f_{\text{adult}}(x, \boldsymbol{\beta}_{\text{adult}}) - f_{\text{ped}}(x, \boldsymbol{\beta}_{\text{ped}}), \quad x \in [a, b]. 
\] 
Our simulation experiences suggest that $S_{D(x;\boldsymbol{\beta}_{\text{ped}})}$ exhibits a decreasing trend with respect to $P(\text{reject } H_0 \mid \text{Data})$.  

Furthermore, as we can see later (in Figure~\ref{fig:3}), a large value of $f_{\text{ped}}(a, \boldsymbol{\beta}_{\text{ped}})$, the value of the pediatric curve at the left endpoint ($x = a = 0$), is indicative of a larger difference at the right endpoint ($x = b = 5$) between the pediatric and adult curves, implying lower similarity. Therefore, an increase in both $S_{D(x;\boldsymbol{\beta}_{\text{ped}})}$ and $f_{\text{ped}}(a, \boldsymbol{\beta}_{\text{ped}})$ corresponds to a decrease in similarity. Hence, we consider $S_{D(x;\boldsymbol{\beta}_{\text{ped}})} \times f_{\text{ped}}(a, \boldsymbol{\beta}_{\text{ped}})$ and map it into interval $[0, 1]$. We define $\eta$ as follows:  
\begin{equation}
    \eta = \frac{S_{D(x;\boldsymbol{\beta}_{\text{ped}})} \times f_{\text{ped}}(a, \boldsymbol{\beta}_{\text{ped}}) - \min\left(S_{D(x;\boldsymbol{\beta}_{\text{ped}})} \times f_{\text{ped}}(a, \boldsymbol{\beta}_{\text{ped}})\right)}{\max\left(S_{D(x;\boldsymbol{\beta}_{\text{ped}})} \times f_{\text{ped}}(a, \boldsymbol{\beta}_{\text{ped}})\right) - \min\left(S_{D(x;\boldsymbol{\beta}_{\text{ped}})} \times f_{\text{ped}}(a, \boldsymbol{\beta}_{\text{ped}})\right)}
    \label{eta_def}
\end{equation}  

In this formulation, $\eta$ is uniquely determined by $\boldsymbol{\beta}_{\text{ped}}$. Additionally, the simulation study presented in Section~5.1 confirms the observed decreasing trend between $\eta$ and $P(\text{reject } H_0 \mid \text{Data})$.

 For an illustration of $\eta$ corresponding to different pediatric curves, consider the following example: Let $\delta=0.2$ in \eqref{eq_2.2} and  $\bfbeta_{\text{adult}}=(-1,1)$, with $[a, b] = [A, B] = [0, 5]$ . In Figure \ref{fig:3}, we display pediatric curves generated under the condition $\max_{x\in[a,b]} D(x) = \delta$, with varying values of $\eta$. As illustrated in Figure~\ref{fig:3}, an increase in $\eta$ reduces the similarity between the pediatric and adult E-R curves at the lower exposure levels, while enhancing their similarity at the higher exposure levels.

	As we discuss above, we employ a decreasing distribution for $\eta$ under $H_0$ and an increasing distribution under $H_1$. These weights are denoted as $\pi_0 (\eta)$ and $\pi_1(\eta)$, respectively. For $\delta$ under $H_1$, we specify the weight $\pi(\delta)$ to be centered around $\epsilon_H/2$. Consequently, the generation of $\boldsymbol{\beta}_{\text{ped}}$  under two hypotheses expressed as: 
 \begin{align}
		\pi^{H_0}(\boldsymbol{\beta}_{\text{ped}})&\propto \int_{0}^1\boldsymbol{\beta}_{\text{ped}}(\epsilon_H;\eta)\pi_{0}(\eta)d\eta \label{sampling00}\\
		\pi^{H_1}(\boldsymbol{\beta}_{\text{ped}})&\propto\int_{0}^1\int_{0}^{\epsilon_{bayes}}\boldsymbol{\beta}_{\text{ped}}(\delta;\eta)\pi_{1}(\eta)\pi(\delta)d\delta d\eta \label{sampling01}
\end{align}
We compute the integrals in equations \eqref{sampling00} and \eqref{sampling01} by discretization. For $\eta$ in \eqref{sampling00} and \eqref{sampling01}, let it be $0.1,0.15,0.25,.\cdots, 0.95$ and $0.1,0.15,0.25,\cdots,0.95,1$, respectively. For $\delta$ in \eqref{sampling01}, we let it be $k\epsilon_H/10,k=0,\cdots,9$. Then, \eqref{sampling00} and \eqref{sampling01} can be approximated as 
	\begin{align}
		\pi^{H_0}(\boldsymbol{\beta}_{\text{ped}})&\propto \sum_{i=1}^{10}\boldsymbol{\beta}_{\text{ped}}(\epsilon_H;\eta_i)\pi_{0}(\eta_i) \label{sampling_1}\\
		\pi^{H_1}(\boldsymbol{\beta}_{\text{ped}})&\propto\sum_{k=0}^9\sum_{i=1}^{11}\boldsymbol{\beta}_{\text{ped}}(k\epsilon_H/10;\eta_i)\pi_{1}(\eta_i)\pi(k\epsilon_H/10) \label{sampling_2}
	\end{align}

Based on \eqref{sampling_1}, to generate $\boldsymbol{\beta}_{\text{ped}}$, we can choose a  $\eta_i\in\bfeta$ with the probability vector $\pi_0(\bfeta)$, then we can find the  $\boldsymbol{\beta}_{\text{ped}}(\epsilon_H;\eta_i)$ corresponding to $\eta_i$ based on \eqref{eta_def}  as $\boldsymbol{\beta}_{\text{ped}}$. Similarly, to generate $\boldsymbol{\beta}_{\text{ped}}$ in \eqref{sampling_2}, we can generate one $\delta$ with the probability vector $\pi(k\epsilon_H/10),k=0,\cdots,9$. Then, we choose a $\eta_i\in\bfeta$ with the probability vector $\pi_1(\bfeta)$. Finally, we can use the $\boldsymbol{\beta}_{\text{ped}}(k\epsilon_H/10;\eta_i)$ as $\boldsymbol{\beta}_{\text{ped}}$. To summarize, our workflow to generate the $\boldsymbol{\beta}_{\text{ped}}$ under two hypotheses can be seen in the Supplementary Web Figure 1.
	\section{Algorithm}
    \subsection{Compute Average Type I error and Power}
	To design a study, we are interested in computing the following quantity, which corresponds to expected type I error rate or power given $H_0$ or $H_1$.                            
\begin{equation}
     \mathbb{E}\left[1(\mathbb{P}\left(\text{ reject }H_0 \mid \mathbf{y}^{(n)},H_0\text{ or }H_1\right) > \epsilon_{\text{bayes}})\right],
    \label{eq:quantitative_measure}
\end{equation}
 where $1(\cdot)$ is the indicator function, $\mathbf{y}^{(n)}$  is the simulated pediatric response data with sample size $n$ and the expectation is with respect to the marginal distribution of $\mathbf{y}^{(n)}$. To compute this, we need $(n, w, \epsilon_{\text{bayes}})$, here  $w$ is the weight for the informative part in REPP, and $\epsilon_{\text{bayes}}$ is the decision threshold. We call $(n, w, \epsilon_{\text{bayes}})$ the design parameter tuple in our proposed extrapolation design

Given the REPP for $\boldsymbol{\beta}_{\text{ped}}$  and data  $\mathbf{y}^{(n)}$,,we can get posterior distribution of $\boldsymbol{\beta}_{\text{ped}}^{(m)},m=1,\cdots,M$,  here $M$ is number of posterior sample. We can then evaluate the following: 

\begin{multline}
    \mathbb{P}\left(\max_{x \in [a, b]}\{f_{\text{adult}}(x, \mathbf{\bfbeta}_{\text{adult}}) - 
    f_{\text{ped}}(x, \mathbf{\bfbeta}_{\text{ped}})\} < \epsilon_H \mid \mathbf{y}^{(n)}, H_0\text{ or }H_1\right) \\
    \approx \frac{1}{M} \sum_{m=1}^M {1}\left\{\max_{x \in [a, b]}\{f_{\text{adult}}(x, 
    \mathbf{\bfbeta}_{\text{adult}}) - f_{\text{ped}}(x, \mathbf{\bfbeta}_{\text{ped}}^{(m)})\} < \epsilon_H\mid \mathbf{y}^{(n)}, H_0\text{ or }H_1\right\},
    \label{eq:approximation_probability}
\end{multline}
We can see that given $H_0$ , \eqref{eq:approximation_probability} refers to type I error rate and given $H_1$, refers to the power.

To evaluate the expected Type I error rate and power for given weight $w$, Bayesian threshold $\epsilon_{\text{bayes}}$, and sample size $n$, we use the following steps:
\begin{enumerate}
    \item \textbf{Sample $\boldsymbol{\beta}_{\text{ped}}$ using \eqref{sampling_1} and \eqref{sampling_2}:} For type I error calculation, we draw  $\boldsymbol{\beta}_{\text{ped}}$  using Equation \eqref{sampling_1}.   For power calculation, for type we draw  $\boldsymbol{\beta}_{\text{ped}}$  using Equation \eqref{sampling_2}. 
    \item \textbf{Generate $\mathbf{y}^{(n)}$:} we first generate the exposure $\mathbf{x}^{(n)}={x_1,…,x_n}$. Assume that $x_i$ are i.i.d and  $x_i$ belongs to $[a, b]$ and $[A, B]-[a, b]$ with equal probability. The data-generating process of $x_i$ is as below:
\begin{enumerate}
    \item[i.] Sample $u\sim U[0,1]$
    \item[ii.] If $u\leq0.5$, we generate $x_i\sim U[a,b]$; if $u>0.5$, we generate $x_i\sim U\{[A,B]-[a,b]\}$;
With $\mathbf{x}^{(n)}$ and $\boldsymbol{\beta}_{\text{ped}}$ in step 1, we can generate $\mathbf{y}^{(n)}$ using Logistic model with Equation \eqref{eq_gene};
\end{enumerate}

    \item \textbf{Obtain the posterior sample of $\boldsymbol{\beta}_{\text{ped}}$:} Apply the Bayesian method to obtain the posterior sample of $\boldsymbol{\beta}_{\text{ped}}$ using REPP. It is implemented using JAGS.
    \item \textbf{Calculate the average type I error and power:} 	Compute the Equation \eqref{eq:approximation_probability} based on the right-hand side formula, denoted as $\mathbb{\hat{P}}\left(\text{ reject }H_0|H_0 \text{ or }H_1 \right)$, and  check whether \eqref{eq:approximation_probability} exceeds $\epsilon_{\text{bayes}}$. Repeat the procedure above for a total of $T$ cycle and calculate the proportion of times \eqref{eq:approximation_probability} exceed $\epsilon_{\text{bayes}}$ over $T$ iterations. That’s
    \begin{align*}
        \text{Average Type I error }&=\frac{1}{T}\sum_{t=1}^T1\left(\mathbb{\hat{P}}\left(\text{ reject }H_0|\mathbf{y}^{(n)}_t,H_0 \right)\right)\\
         \text{Average power }&=\frac{1}{T}\sum_{t=1}^T1\left(\mathbb{\hat{P}}\left(\text{ reject }H_0|\mathbf{y}^{(n)}_t,H_1 \right)\right)
    \end{align*}
\end{enumerate}
\subsection{Determining the Optimal Parameters $(n, w, \epsilon_{\text{bayes}})$}

Consider the design scenario where both the type I error and type II error are constrained by predefined thresholds $\alpha$ and $\beta$, respectively. The design parameter tuple $(n, w, \epsilon_{\text{bayes}})$ is defined as a \emph{qualified design tuple} if it satisfies the following conditions:  
\begin{itemize}
    \item The type I error does not exceed $\alpha$.  
    \item The power is no less than $1-\beta$.  
\end{itemize}

However, due to simulation variability, it is observed that when the difference between the calculated average type I error and $\alpha$ is within 0.05, or when the difference between the calculated average power and $1-\beta$ is within 0.05, the qualified parameter tuples may depend on $T$, which is the number of simulation iterations.  

To mitigate this issue, we introduce the concept of \emph{admissible candidate design tuples}, where the parameters $(n, w, \epsilon_{\text{bayes}})$ satisfy the following criteria:  

\begin{enumerate}
    \item[(1)] For the observed type I error, $\alpha^{\ast}$, and power, $1-\beta^{\ast}$, at most one constraint may be violated. Specifically, the following conditions may occur:
    \begin{itemize}
        \item $\alpha^{\ast} \leq \alpha$ and $1-\beta^{\ast} < 1-\beta$, or  
        \item $\alpha^{\ast} > \alpha$ and $1-\beta^{\ast} \geq 1-\beta$.  
    \end{itemize}
    
    \item[(2)] For any violated constraint, the deviation must be within an acceptable margin of 0.05. For instance, if the power constraint is violated, the absolute difference between $1-\beta^{\ast}$ and the prespecified $1-\beta$ must be within 0.05.
\end{enumerate}

Upon identifying the set of \emph{admissible} tuples, one common approach for selecting the optimal design is to choose the tuple with the smallest sample size, $n$. However, in scenarios where multiple admissible parameter combinations exist, additional considerations may arise beyond sample size. For example, consider the following two admissible tuples:  
\begin{itemize}
    \item $(40, 0.2, 0.99)$ with type I error and power values of $(0.20, 0.67)$, and  
    \item $(45, 0.1, 0.95)$ with type I error and power values of $(0.15, 0.70)$.  
\end{itemize}
While the first tuple requires a smaller sample size, some practitioners may prefer the second option, accepting a larger sample size in exchange for higher power and reduced reliance on borrowing information from adult data (indicated by a lower value of $w$).  

Therefore, the selection of optimal parameters should be made in consultation with health authorities, considering multiple factors such as sample size, statistical power, and the degree of adult data incorporation. This approach ensures a balanced and context-sensitive decision-making process.

We propose a measure to evaluate the performance of controlling Type I and Type II errors for each admissible tuple, facilitating the selection of an optimal tuple based on its ability to maintain Type I error and Type II error control . For Type I error control, under the null hypothesis ($H_0$), we select a value of $\boldsymbol{\beta}_{\text{ped}}$ from the set of $\boldsymbol{\beta}_{\text{ped}}$ corresponding to $\eta$ ranging from 0.1 to 0.95, as defined by Equation \eqref{sampling_1}.  The Type I error for each selected $\boldsymbol{\beta}_{\text{ped}}$ is then computed, and the proportion of Type I error $\leq\alpha$  can be calculated. Using the similar procedure, under alternative hypothesis ($H_1$), we calculate the power of each $\boldsymbol{\beta}_{\text{ped}}$  according to Equations \eqref{sampling_2}. Then, we can compute the proportion of  power $\geq1-\beta$. This proportion serves as an indicator of the stability of Type I and Type II error control, with higher proportions reflecting greater stability. To select the optimal design parameters $(n, w, \epsilon_{\text{bayes}})$, we prioritize tuples with high proportions, for example, those exceeding 0.8 for both Type I error control and power. It is worth noting that evaluating the stability of $(n, w, \epsilon_{\text{bayes}})$ is not mandatory but serves as a valuable assessment of the design's capacity to maintain Type I error control and enhance statistical power.  
 
 \section{Pediatric Approval Case Example}
Darunavir is a protease inhibitor (PI) for HIV/AIDS treatment, inhibiting the HIV-1 protease essential for viral replication by preventing polyprotein cleavage, thereby reducing viral load and improving immune function \citep{FDA:label:2008}.

In June 2006, the FDA approved darunavir based on the TITAN (TMC114/r In Treatment-experienced patients Naive to lopinavir) phase III trial and two phase IIb trials, demonstrating significant viral load reduction and CD4+ count increases in treatment-experienced adults.

In December 2008, FDA approval extended to pediatric patients ($\geq$3 years) based on the DELPHI (Darunavir Evaluation in Pediatric HIV-1-Infected treatment-experienced patients) phase II trial, which confirmed the pharmacokinetics, safety, and efficacy of ritonavir-boosted darunavir in reducing viral load in children, consistent with adult outcomes.

The DELPHI trial used the inhibitory quotient (IQ), the ratio of steady-state drug exposure ($C_{0h}$) to the IC50 value, to measure response. The virologic response was defined as a plasma viral load reduction from baseline of $\geq$1.0 $\log_{10}$ by Week 24—a binary outcome.

\begin{figure}
    \centering
    \begin{subfigure}[b]{0.48\textwidth}
        \centering
        \includegraphics[width=\textwidth]{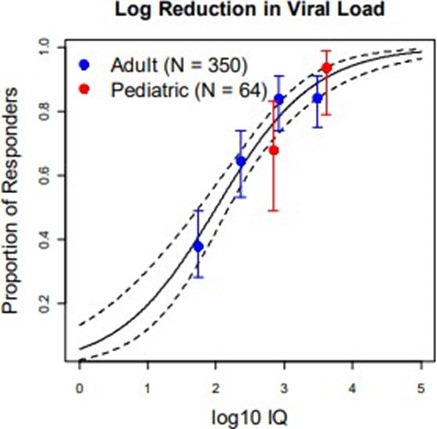}
        \caption{The exposure response curve.}
        \label{fig:4}
    \end{subfigure}
    \hfill
    \begin{subfigure}[b]{0.6\textwidth}
        \centering
        \includegraphics[width=\textwidth]{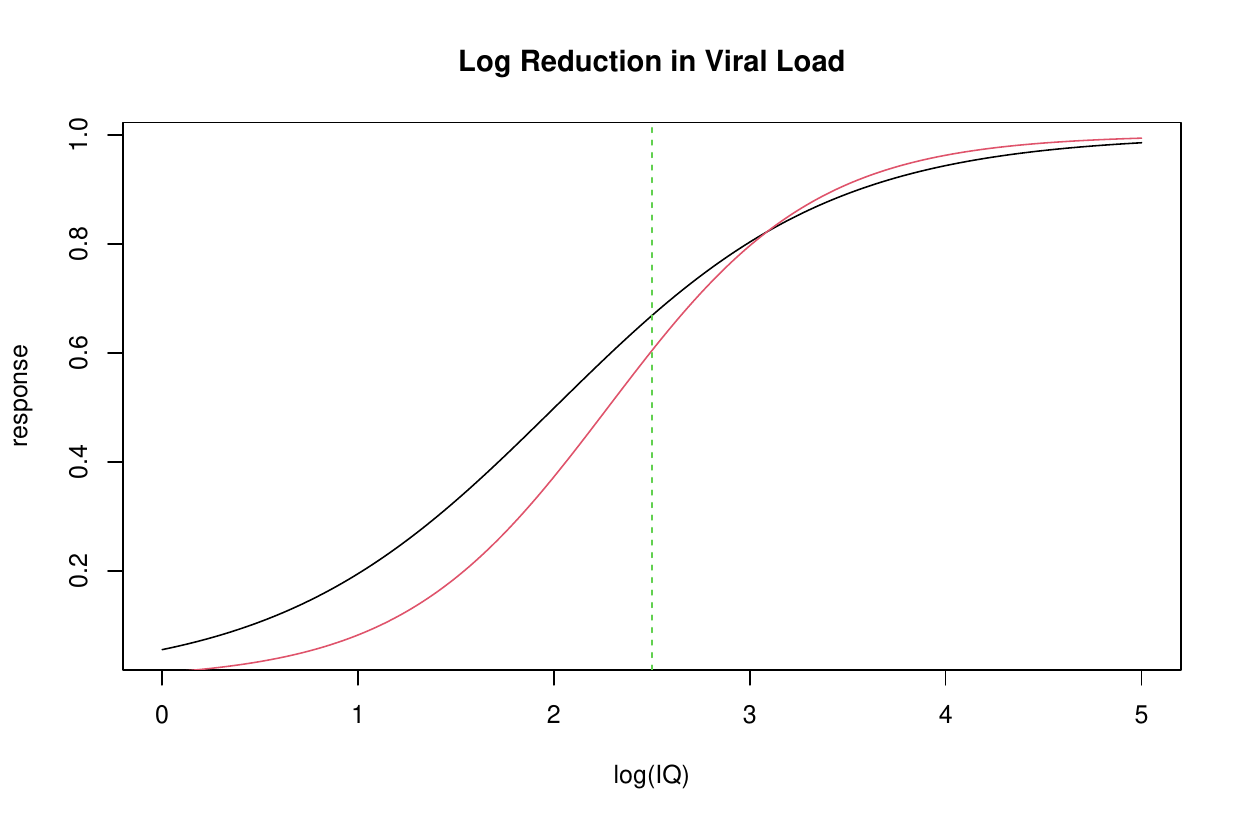}
        \caption{Black line: Adult E-R curve; Red line: Reconstructed pediatric E-R curve.}
        \label{cons}
    \end{subfigure}
    \caption{Exposure-response curves for adult and pediatric populations in different settings.}
\end{figure}

Figure \ref{fig:4} from \cite{FDA:2008} shows the exposure-response curve, with $\log_{10}$ IQ on the x-axis and proportion of responders on the y-axis. Proportions were calculated using all patients assessed at Week 24, with responders meeting the virologic criteria. Blue and red dots with vertical lines represent quantile plots for adult and pediatric datasets, respectively. Adults were grouped into four quartiles, while pediatric data used two quantiles due to smaller sample size. The logistic regression curve from the adult cohort serves as a reference for comparing pediatric responses.

Data from Figure \ref{fig:4} was extracted using PlotDigitizer software, yielding logistic regression coefficients for the adult population as $\bfbeta_{\text{adult}}= (-2.83, 1.41)$. Our analysis has two primary objectives:
\begin{enumerate}
    \item[1.] \textbf{Design:} Assuming we only know the adult data, how to use the above proposed design to design an extrapolation study for pediatric.
    \item[2.] \textbf{Trial Data Analysis:} With both adult and pediatric data like showing in this example, to determine if the similarity between adult and pediatric populations can be established using our proposed criteria.
\end{enumerate}
\textbf{Design:} To design a pediatric trial, given the knowledge learned from adult trials, the exposure range was set to $[0,5]$, and the interested exposure range of pediatric was $[a, b] = [2.5, 5]$, since in this range the treatment shows efficacy on adult patients. We set the threshold $\epsilon_H=0.2$ which was derived from the published adult and pediatric curves. The number of simulation runs was set at $T = 1000$, with Type I and Type II error rates targeted to be $\alpha=0.2$ and $\beta = 0.3$.

 To use the REPP, we selected the points $\bfx=(2.5,3.125,4.375)$, which correspond to the minimum, 25th percentile, and 75th percentile of the interval $[2.5, 5]$. To quantify potential differences between pediatric and adult exposure-response (E-R) curves, it is assumed that the absolute mean of the deviation at each point in $\bfx = (2.5, 3.125, 4.375)$ corresponds to 5\% of the respective log-odds ratio of the adult E-R curve. Additionally, based on evidence suggesting that pediatric responses may exceed adult responses at higher exposure levels, it is hypothesized that the response rate in pediatric patients is lower than that in adults at $x = 2.5$, but higher at $x = 3.125$ and $4.375$. Accordingly, the sign of the mean at $\bfx$ is negative, positive, and positive, respectively. A sensitivity analysis of the mean was conducted (see Section 5.2), revealing that the optimal choice remains relatively stable with respect to variations in the mean. For standard deviation, we assume $\sigma_B(x)=0.01$ at these points.    
 
For the weights assigned to $\bfeta$, as previously noted which is used to generate $\boldsymbol{\beta}_{\text{ped}}$, the weights decrease under $H_0$ and increase under $H_1$ as discussed before. Specifically, we set $\pi_0(\bfeta)=(8, 4, 2, 2, 2, 1, 1, 1, 1, 1)$ under $H_0$ and $\pi_1 (\bfeta) = (1, 1, 1, 1, 1, 1, 2, 2, 2, 4, 8)$ under $H_1$. For the weight of  $\delta$, we set $\pi(\bfdelta) = (1, 2, 4, 8, 4, 2, 1, 1, 1, 1)$, which centers $\delta$ at $0.3\epsilon_H$. These weights,  $\pi_0(\bfeta)$, $\pi_1(\bfeta)$  and $\pi(\bfdelta)$,  were selected so that the largest weight is eight times as the smallest, decreasing by a factor of 2. Sensitivity analyses have confirmed that the results are robust to these weight choices (refer to Section 5.2).

For $(n, w, \epsilon_{\text{bayes}})$, we set sample size, $n$, to be $\{40,45,50,55,60,64\}$, the weight of adult information, $w$,  to be $\{0.1,0.2,0.3,0.4,0.5\}$ and Bayesian threshold $\epsilon_{\text{bayes}}$ to be $\{0.8,0.85,0.9,0.95,0.99\}$. Table \ref{table:all_result} presents the combination of design parameter tuple.  It shows that the probability of rejecting $H_0$ generally increases with $w$, which is expected due to the increased reliance on adult information. The design tuples that satisfy both type I error less than 0.2 and type II error less than 0.3 are marked in red. For the candidate tuples that only one constraint are not satisfied with either power is between 0.65 and 0.7 or type I error is between 0.2 and 0.25 are highlighted in blue. For the blue tuples, for example, when $n=55$, we can see $(w, \epsilon_{\text{bayes}})$ could be (0.1, 0.95), (0.2, 0.95) and (0.2,0.99), three of which are all satisfy the admissible criteria.

    From Table \ref{table:all_result}, we observe the following: 
    \begin{enumerate}
        \item[1.] For a given sample size, both Type I error and power increase with $w$, which means that if the study borrow more adult information, the there is a tradeoff between type I error and power.
        \item[2.] With the increase of the sample size n, we have more qualified design tuples.
    \end{enumerate}
Based on these observations, The design parameter tuple can be chosen from the red or blue ones shown in Table \ref{table:all_result}. For example, a pediatric trial with a sample size 45 is an acceptable design, allowing 10\% ($w=0.1$) borrowing from adult trial and declaring the similarity between two E-R curves when posterior probability, $P ( \text{reject } H_0|Data)$, exceeds 95\% ( $\epsilon_{\text{bayes}}=0.95$) . The design is acceptable because the type I error is controlled at 0.165 and power at 70.5\% when the maximum differences allowed are 0.2 between pediatric and adult E-R curves. Also, if the study cannot afford sample size of 45, another option is to choose design tuple $(n, w, \epsilon_{\text{bayes}})$  as $(40,0.1,0.95)$. In this design, the type I error rate can be controlled at 13.5\% and power is 65\%.  The final decision will be made case-by-case and the discussion will be with regulatory bodies.

Since there are multiple design parameter tuples to choose, to aid in finding the optimal design parameter tuple $(n, w, \epsilon_{\text{bayes}})$, we evaluate the performance to control type I and type II error for each admissible tuples . As we mentioned before, the evaluation is to test whether admissible tuples  are robust to control the type I error and keep the power. To do so, we need to calculate the type I and power with different fixed $\boldsymbol{\beta}_{\text{ped}}$. Under $H_0$, we choose the $\boldsymbol{\beta}_{\text{ped}}$  corresponding to $\eta=0.1, 0.15,\cdots, 0.95$. Under $H_1$, we fix $\delta$ to be 0.06 and choose the $\boldsymbol{\beta}_{\text{ped}}$ corresponding to $\eta= 0.1, 0.15,\cdots, 0.85, 0.95, 1$.  We compute Type I error and power for each coefficient under $H_0$ and $H_1$, and determine the proportions that satisfy the criteria (Type I error $\leq\alpha$ and power $\geq 1-\beta$). We mark the proportions that are no less than 0.8 as red in Table \ref{table:2}. From the result in Table \ref{table:2}, if we choose $(n, w, \epsilon_{\text{bayes}})$ to be $(40, 0.1, 0.95)$, the desired power 0.7 can be achieved 36\% of the times when the ER curve difference between pediatric and adults are smaller than 0.06, thus not stable enough to establish similarity. Hence, if the sample size can be increased to 50, using design parameters $(n, w, \epsilon_{\text{bayes}})$ of $(50, 0.1, 0.95)$ is a better balance to achieve stable performance.

\textbf{Trial Data Analysis:} As indicated earlier, darunavir approval in the pediatric population was based on a sample size of 64. To compare our design identified by the method to the actual trial, from Table \ref{table:all_result}, we can select either $w=0.1$, $\epsilon_{\text{bayes}}=0.95$ or $w=0.2$, $\epsilon_{\text{bayes}}=0.95$. However, Table \ref{table:2} indicates that $w=0.1$, $\epsilon_{\text{bayes}}=0.95$ demonstrates more stable performance than $w=0.2$, $\epsilon_{\text{bayes}}=0.95$. Additionally, if the preference by regulatory agency is to use minimal adult information, the design parameters of $w=0.1$, $\epsilon_{\text{bayes}}=0.95$ would be the choice. 

For the evaluation of the curves observed in the actual trial, the pediatric exposure-response curve was reconstructed to align with two key reference points. The first reference point was defined by the x-coordinate of the first vertical red line in Figure \ref{fig:4}, with the corresponding y-coordinate set as the midpoint between the median response and the upper bound of its 95\% confidence interval. The second reference point was the median response of the second pediatric quantile group, represented by the red dot in Figure \ref{fig:4}. The resulting logistic curve is presented in Figure \ref{cons}. The maximum distance between the adult and pediatric curves within the range $x \in [2.5, 5]$ was 0.0643, with pediatric coefficients estimated as $\beta_{\text{ped}} = (-4.293, 1.886)$. The covariate, $\log_{10}$(IQ), obtained from the clinical pharmacology review \citep{FDA:2008}, was used to generate the test dataset. Based on our analysis, we accepted $H_1$, indicating similarity between the pediatric and adult exposure-response curves.

	\begin{table}[htbp]
    \centering
    \resizebox{\textwidth}{!}{\begin{tabular}{|l|l|l|l|l|l|l|}
        \hline
        sample size & \diagbox{$w$}{$\epsilon_{\text{bayes}}$} & 0.8 & 0.85 & 0.9 & 0.95 & 0.99 \\ \hline
        \multirow{5}{*}{$n=40$}  
        & 0.1 & (0.40,0.850) & (0.305,0.815) & (0.240,0.750) & {\color{blue}(0.135,0.650)} & (0.025,0.375) \\ \cline{2-7}
        & 0.2 & (0.645,0.920) & (0.535,0.890) & (0.440,0.850) & (0.305,0.775) & (0.080,0.515) \\ \cline{2-7}
        & 0.3 & (0.665,0.950) & (0.610,0.930) & (0.500,0.875) & (0.340,0.790) & (0.155,0.585) \\ \cline{2-7}
        & 0.4 & (0.725,0.955) & (0.660,0.940) & (0.565,0.895) & (0.410,0.820) & (0.120,0.580) \\ \cline{2-7}
        & 0.5 & (0.740,0.965) & (0.665,0.940) & (0.560,0.935) & (0.365,0.855) & (0.100,0.610) \\ \hline \hline

        \multirow{5}{*}{$n=45$}  
        & 0.1 & (0.455,0.880) & (0.345,0.865) & (0.26,0.820) & {\color{red}(0.165,0.705)} & (0.050,0.430) \\ \cline{2-7}
        & 0.2 & (0.570,0.935) & (0.470,0.910) & (0.36,0.825) & (0.245,0.760) & (0.090,0.525) \\ \cline{2-7}
        & 0.3 & (0.625,0.930) & (0.555,0.910) & (0.44,0.885) & (0.285,0.830) & (0.085,0.545) \\ \cline{2-7}
        & 0.4 & (0.710,0.920) & (0.615,0.900) & (0.49,0.875) & (0.335,0.835) & (0.100,0.630) \\ \cline{2-7}
        & 0.5 & (0.755,0.965) & (0.705,0.935) & (0.60,0.910) & (0.435,0.835) & (0.155,0.600) \\ \hline \hline

        \multirow{5}{*}{$n=50$}  
        & 0.1 & (0.425,0.900) & (0.340,0.860) & (0.265,0.795) & {\color{red}(0.150,0.705)} & (0.040,0.475) \\ \cline{2-7}
        & 0.2 & (0.530,0.915) & (0.490,0.890) & (0.39,0.845) & (0.260,0.785) & (0.070,0.510) \\ \cline{2-7}
        & 0.3 & (0.645,0.930) & (0.560,0.915) & (0.47,0.890) & (0.290,0.835) & (0.105,0.630) \\ \cline{2-7}
        & 0.4 & (0.615,0.935) & (0.530,0.925) & (0.44,0.910) & (0.355,0.840) & {\color{blue}(0.125,0.655)} \\ \cline{2-7}
        & 0.5 & (0.760,0.970) & (0.655,0.960) & (0.520,0.925) & (0.360,0.890) & (0.1,0.640) \\ \hline \hline

        \multirow{5}{*}{$n=55$}  
        & 0.1 & (0.510,0.880) & (0.395,0.845) & (0.330,0.770) & {\color{blue}(0.175,0.690)} & (0.015,0.425) \\ \cline{2-7}
        & 0.2 & (0.540,0.900) & (0.505,0.880) & (0.430,0.845) & (0.245,0.755) & (0.065,0.535) \\ \cline{2-7}
        & 0.3 & (0.690,0.955) & (0.640,0.950) & (0.535,0.920) & (0.350,0.845) & {\color{blue}(0.140,0.675)} \\ \cline{2-7}
        & 0.4 & (0.615,0.945) & (0.550,0.935) & (0.450,0.910) & (0.345,0.830) & (0.110,0.605) \\ \cline{2-7}
        & 0.5 & (0.700,0.940) & (0.620,0.920) & (0.555,0.905) & (0.375,0.870) & (0.105,0.645) \\ \hline \hline

        \multirow{5}{*}{$n=60$}  
        & 0.1 & (0.455,0.875) & (0.415,0.845) & (0.335,0.795) & {\color{red}(0.20,0.72)} & (0.040,0.490) \\ \cline{2-7}
        & 0.2 & (0.520,0.925) & (0.455,0.890) & (0.365,0.84) & (0.240,0.750) & (0.070,0.55) \\ \cline{2-7}
        & 0.3 & (0.585,0.910) & (0.485,0.885) & (0.400,0.87) & (0.255,0.800) & (0.09,0.620) \\ \cline{2-7}
        & 0.4 & (0.695,0.950) & (0.615,0.945) & (0.545,0.940) & (0.365,0.885) & {\color{red}(0.085,0.70)} \\ \cline{2-7}
        & 0.5 & (0.685,0.950) & (0.610,0.925) & (0.530,0.89) & (0.325,0.850) & (0.10,0.65) \\ \hline \hline

        \multirow{5}{*}{$n=64$}  
        & 0.1 & (0.460,0.875) & (0.379,0.850) & (0.290,0.810) & {\color{red}(0.20,0.740)} & (0.040,0.525) \\ \cline{2-7}
        & 0.2 & (0.465,0.930) & (0.400,0.925) & (0.320,0.895) & {\color{red}(0.195,0.805)} & (0.050,0.575) \\ \cline{2-7}
        & 0.3 & (0.565,0.910) & (0.460,0.895) & (0.420,0.825) & (0.285,0.775) & (0.100,0.585) \\ \cline{2-7}
        & 0.4 & (0.740,0.935) & (0.675,0.925) & (0.515,0.895) & (0.340,0.840) & (0.055,0.640) \\ \cline{2-7}
        & 0.5 & (0.695,0.965) & (0.610,0.965) & (0.525,0.950) & (0.350,0.875) & {\color{blue}(0.085,0.685)} \\ \hline
    \end{tabular}
    }
    \caption{First number in the bracket is type I error. Second number is the power = 1- Type II error. The red numbers are the designs satisfying the type I and II error constraint (Type I $\leq$ 0.2, Type II error $\leq$ 0.3). Blue ones are the potential candidate designs. The results are based on $\epsilon_H=0.2$.}
    \label{table:all_result}
\end{table}

%\begin{table}[htbp]
%		\centering
%		\begin{tabular}{|l|l|l|}
%  \hline
%  &$\bfbeta_{\text{ped}}$&max value of difference\\ \hline
%\multirow{5}{*}{Power} &$(-2.86,1.35)$&0.0424\\ \cline{2-3}
% &$(-2.97,1.39)$&0.0454\\ \cline{2-3}
% &$(-3.02,1.39)$&0.0591\\ \cline{2-3}
% &$(-3.02,1.38)$&0.0620\\ \cline{2-3}
% &$(-2.88,1.31)$&0.0718\\ \hline
%\multirow{5}{*}{Type I} &$(-2.98,1.14)$&0.200\\ \cline{2-3}
% &$(-3.24,1.23)$&0.208\\ \cline{2-3}
% &$(-3.20,1.22)$&0.209\\\cline{2-3}
 %&$(-3.11,1.17)$&0.212\\\cline{2-3}
 %&$(-3.27,1.23)$&0.213\\\hline
%		\end{tabular}
%		\caption{Coefficients Table.}
%		\label{table:coef}
%\end{table}
\begin{table}
\centering
\caption{Proportion that satisfies the Type I and II error constraints. First number in the bracket is the proportion with respect to type I error. Second number is the proportion with respect to power. Blank cell means the result is not available. The pairs in red color are the pairs satisfying the constraints that the two proportions are no less than 0.8.}
\label{table:2}
\begin{tabular}{|c|c|c|c|c|c|}
\hline
\multirow{2}{*}{sample size} & \multicolumn{5}{c|}{$(w, \epsilon_{\text{bayes}})$} \\
\cline{2-6}
 & (0.1,0.95) & (0.2,0.95) & (0.3,0.99) & (0.4,0.99) & (0.5,0.99) \\
\hline
40 & (0.8,0.36) & & & & \\
\hline
45 & (1,0.54) & & & & \\
\hline
50 & \textcolor{red}{(0.9,1)} & & & (0,1) & \\
\hline
55 & (0.2,1) & & (0.7,1) & & \\
\hline
60 & (0.7,1) & & & (0.6,1) & \\
\hline
64 & \textcolor{red}{(0.8,1)} & (0.1,1) & & & (0.6,1) \\
\hline
\end{tabular}
\end{table}

 	\section{Simulation Studies}
	\subsection{Trend between $\eta$ and $P(\text{ reject }H_0|Data)$}
\begin{figure}
		\includegraphics[scale=1]{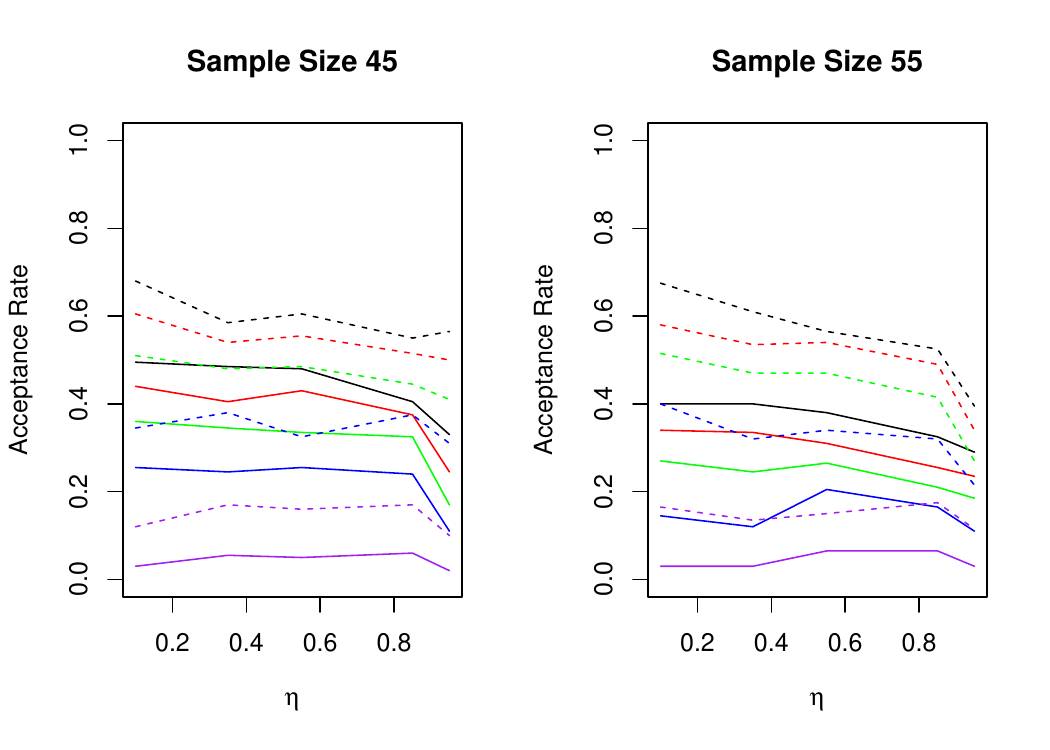}
		\caption{The acceptance rate v.s. $\eta$ within sample size 45 and 55. Different types of line means different weight of adult information. Solid: $w=0.1$; dashed: $w=0.3$. Different colors means different $\epsilon_{\text{bayes}}$. Black: $0.8$; red: $0.85$; green: $0.90$; blue: $0.95$; purple: 0.99. }
		\label{trend}
	\end{figure}

In this study, we investigate the trend between $\eta$ and $P(\text{ reject }H_0|Data)$. By employing logistic regression models calibrated using empirical data, the adult population model is parameterized as  $\boldsymbol{\beta}_{\text{adult}} = (-2.83, 1.41)$. The interval of interest is defined as $[a, b] = [2.5, 5]$, with $\epsilon_H=0.2$  and maximum difference $\delta = 0.2$ in Equation \eqref{eq_2.2}, and hence $P(\text{ reject }H_0|Data)$ is Type I error rate. Sample sizes are selected from $n=(45, 55)$. To simplify the analysis, the selection of $w$, and $\eta$ is made tractable by choosing $w$ from the set $(0.1, 0.3)$ and $\eta$ from $(0.1, 0.35, 0.55, 0.85, 0.95)$. For each specified value of $\eta$  within the specified set, corresponding parameters $\boldsymbol{\beta}_{\text{ped}}$ are determined, and a simulation study is conducted over $T = 300$ trials. The outcomes, depicted in Figure \ref{trend}, substantiate the statement in Section 2.2 that acceptance rate define in $P(\text{ reject }H_0|Data)$  decreases with $\eta$.
 
	\subsection{Sensitivity Analysis}
Given the numerous parameters to specify in our model, we conducted a sensitivity analysis to test the stability of our parameter choices across different scenarios:
 \begin{enumerate}
     \item[(1).] \textbf{Changing Weights for $\bfeta$:} We altered the weights for  $\pi_0(\bfeta)$  and $\pi_1(\bfeta)$  to be (4,2,2,1,1,1,1,1,1,1) and (1,1,1,1,1,1,1,1,2,2,4), respectively. This case represents an increased probability of selecting the elements of $\bfeta$ with weight 1, thereby expanding the search space for $\bfeta$. 
     \item[(2).] \textbf{Modified Probability Ratios:} We adjusted the ratio of the probability of $x\in [a,b]$ to $x\in [A,B]-[a,b]$ by doubling it,  aiming to better reflect real-world conditions, namely, the exposure in pediatric population will concentrate on high level.
     \item[(3).] \textbf{Magnitude of Difference in REPP for $\bfbeta_{\text{ped}}$:} We assume that the mean of deviation at each point in $\bfx=(2.5,3.125,4.375)$ is 10\% or 15\% of the corresponding adult E-R log odds ratio. 
 \end{enumerate}
Due to space limitations, the detailed results and findings are presented in the supplementary material. The final analysis indicates that candidate selections are consistent for $\epsilon_{\text{bayes}} = 0.95$, demonstrating that the final choice of $(n, w, \epsilon_{\text{bayes}})$ remains unchanged. These results suggest that the proposed method exhibits stability in selecting $(n, w, \epsilon_{\text{bayes}})$.

 \section{Conclusion}

Pediatric drug development faces methodological challenges due to ethical constraints and limited patient availability. As a result, regulatory agencies advocate for pediatric extrapolation, which relies on existing adult data to inform pediatric dose selection and efficacy assessment. However, current strategies lack a standardized, quantitative method for evaluating whole-curve exposure–response (E–R) similarity. This study introduces the Pediatric Bayesian Extrapolation Design (PED), a statistical framework that rigorously evaluates E–R similarity while maintaining type I and II error control.

A key feature of PED is the introduction of three methodological components. First, the Robust Elicited Points Prior (REPP) flexibly incorporates expert knowledge into the Bayesian framework. Second, E–R similarity is quantified using the maximum deviation metric, providing a global measure for comparing adult and pediatric E–R profiles. Third, the parameter $\eta$ facilitates efficient coefficient selection by balancing prior adult information and ensuring statistical robustness. These components are integrated through a grid search strategy to identify optimal parameter configurations across sample size, prior weight, and posterior threshold, thereby enhancing the stability of the study design.

We also propose a stability measure for type I and II error control across design parameters. For each admissible tuple $(n, w, \epsilon_{\text{bayes}})$, the proportion of cases where type I error remains below $\alpha$ and power exceeds $1-\beta$ is evaluated. Tuples with high proportions (e.g., $>0.8$) are prioritized, aiding robust design selection. While optional, this assessment enhances reliability of parameters choice.

The practical utility of PED is demonstrated in both prospective and retrospective pediatric drug development scenarios. Sensitivity analyses confirm robustness under varying model specifications. PED is flexible and extendable to alternative similarity metrics (e.g., $L_2$ distance) and more complex dose-response models. Open-source implementation is provided via GitHub (https://github.com/zhongheng-Biostatistics/Pediatric-Bayesian-Extrapolation-Design) to facilitate broader adoption in regulatory and clinical trial planning.

Despite its strengths, PED has limitations. It is currently tailored to logistic E–R curves, with extensions to continuous endpoints as future work. Incorporating hierarchical Bayesian models could improve information borrowing across subpopulations. Further integration with adaptive trial designs may enhance flexibility, especially where pediatric data is limited.

In summary, PED provides a structured, reproducible framework for Bayesian pediatric extrapolation, balancing prior adult data with rigorous statistical control. This approach can streamline regulatory approval, optimize trial designs, and improve access to effective pediatric therapies.

%  The \backmatter command formats the subsequent headings so that they
%  are in the journal style.  Please keep this command in your document
%  in this position, right after the final section of the main part of 
%  the paper and right before the Acknowledgements, Supporting Information (Supplementary %  Materials),   and References sections. 

\backmatter

%  This section is optional.  Here is where you will want to cite
%  grants, people who helped with the paper, etc.  But keep it short!

%\section*{Acknowledgements}

%The authors thank Professor A. Sen for some helpful suggestions,
%Dr C. R. Rangarajan for a critical reading of the original version of the
%paper, and an anonymous referee for very useful comments that improved
%the presentation of the paper.\vspace*{-8pt}

%  Here, we create the bibliographic entries manually, following the
%  journal style.  If you use this method or use natbib, PLEASE PAY
%  CAREFUL ATTENTION TO THE BIBLIOGRAPHIC STYLE IN A RECENT ISSUE OF
%  THE JOURNAL AND FOLLOW IT!  Failure to follow stylistic conventions
%  just lengthens the time spend copyediting your paper and hence its
%  position in the publication queue should it be accepted.

%  We greatly prefer that you incorporate the references for your
%  article into the body of the article as we have done here 
%  (you can use natbib or not as you choose) than use BiBTeX,
%  so that your article is self-contained in one file.
%  If you do use BiBTeX, please use the .bst file that comes with 
%  the distribution.  In this case, replace the thebibliography
%  environment below by 
%
\bibliographystyle{plainnat}
\bibliography{mybibilo.bib}

\begin{thebibliography}{14}
\providecommand{\natexlab}[1]{#1}
\providecommand{\url}[1]{\texttt{#1}}
\expandafter\ifx\csname urlstyle\endcsname\relax
  \providecommand{\doi}[1]{doi: #1}\else
  \providecommand{\doi}{doi: \begingroup \urlstyle{rm}\Url}\fi

\bibitem[Chen and Ibrahim(2000)]{Chen:Ibra:2000}
MingHui Chen and Joseph~G. Ibrahim.
\newblock Power prior distributions for regression models.
\newblock \emph{Statistical Science}, 15:\penalty0 46--60, 2000.

\bibitem[Dette et~al.(2018)Dette, M{\"o}llenhoff, Volgushev, and Bretz]{Dette:2018}
Holger Dette, Kathrin M{\"o}llenhoff, Stanislav Volgushev, and Frank Bretz.
\newblock Equivalence of regression curves.
\newblock \emph{Journal of the American Statistical Association}, 113:\penalty0 711--729, 2018.

\bibitem[Dette et~al.(2024)Dette, Koletzko, and Bretz]{Dette:2024}
Holger Dette, Lukas Koletzko, and Frank Bretz.
\newblock Testing for similarity of dose response in multi-regional clinical trials, 2024.
\newblock URL \url{https://arxiv.org/pdf/2404.17682}.

\bibitem[E11(a)(2022)]{ICHE11A:2022}
E11(a).
\newblock Ich guideline e11a on pediatric extrapolation (ich) (2022, april 6)., 2022.
\newblock URL \url{https://www.ema.europa.eu/en/documents/scientific-guideline/draft-ich-guideline-e11a-pediatric-extrapolation-step-2b_en.pdf}.

\bibitem[FDA(2008{\natexlab{a}})]{FDA:2008}
FDA.
\newblock Office of clinical pharmacology review, 2008{\natexlab{a}}.
\newblock URL \url{https://www.fda.gov/media/75532/download}.

\bibitem[FDA(2008{\natexlab{b}})]{FDA:label:2008}
FDA.
\newblock Fda-approved drugs, 2008{\natexlab{b}}.
\newblock URL \url{https://www.accessdata.fda.gov/drugsatfda_docs/label/2008/021976s009lbl.pdf}.

\bibitem[Hobbs et~al.(2011)Hobbs, Carlin, Mandrekar, and Sargent]{Hobbs:2011}
Brian~P. Hobbs, Bradley~P. Carlin, Sumithra~J. Mandrekar, and Daniel~J. Sargent.
\newblock Hierarchical commensurate and power prior models for adaptive incorporation of historical information in clinical trials.
\newblock \emph{Biometrics}, 67:\penalty0 1047--1056, 2011.

\bibitem[Mulberg et~al.(2022)Mulberg, Conklin, Croft, and for Children Pediatric Extrapolation Working~Group]{Mulberg:2022}
Andrew~E. Mulberg, Laurie~S. Conklin, Nicholas~M. Croft, and I-ACT for Children Pediatric Extrapolation Working~Group.
\newblock Pediatric extrapolation of adult efficacy to children is critical for efficient and successful drug development.
\newblock \emph{Gastroenterology}, 163:\penalty0 77--83, 2022.

\bibitem[Oakley and O'Hagan(2023)]{Oakley:2023}
Jeremy Oakley and Tony O'Hagan.
\newblock Shelf: The sheffield elicitation framework (version 4), 2023.
\newblock URL \url{https://shelf.sites.sheffield.ac.uk/}.

\bibitem[Park and Casella(2008)]{Park:Casella:2008}
Trevor Park and George Casella.
\newblock The bayesian lasso.
\newblock \emph{Journal of the American Statistical Association}, 103:\penalty0 681--686, 2008.

\bibitem[Schmidli et~al.(2014)Schmidli, Gsteiger, Roychoudhury, {\'O}Hagan, Spiegelhalter, and Neuenschwander]{Schmidli:2011}
Heinz Schmidli, Sandro Gsteiger, Satrajit Roychoudhury, Anthony {\'O}Hagan, David Spiegelhalter, and Beat Neuenschwander.
\newblock Robust meta-analytic-predictive priors in clinical trials with historical control information.
\newblock \emph{Biometrics}, 70:\penalty0 1023--1032, 2014.

\bibitem[Ye et~al.(2023)Ye, Zhang, Strimenopoulou, Zhao, Pan, Shabbout, and Gamalo]{Ye:2023}
Jingjing Ye, Vickie Zhang, Foteini Strimenopoulou, Yihua Zhao, Haitao Pan, Mayadah Shabbout, and Margaret Gamalo.
\newblock Recent use of pediatric extrapolation in pediatric drug development in us.
\newblock \emph{Journal of Biopharmaceutical Statistics}, 33:\penalty0 681–695, 2023.

\bibitem[Yu and Moyeed(2001)]{Yu:Moye:2001}
Keming Yu and Rana~A. Moyeed.
\newblock Bayesian quantile regression.
\newblock \emph{Statistics and Probability Letters}, 54:\penalty0 437--447, 2001.

\bibitem[Zhang et~al.(2021)Zhang, Travis, Rothwell, Jay, Jahidur, Zhang, Crentsil, Altepeter, Lee, Burckart, Ganley, and Wang]{Zhang:2021}
Qunshu Zhang, James Travis, Rebecca Rothwell, Christopher~E. Jay, Rashid Jahidur, Yifei Zhang, Victor Crentsil, Tara Altepeter, Jessica~J. Lee, Gilbert~J. Burckart, Charles Ganley, and Jian Wang.
\newblock Applying the noninferiority paradigm to assess exposure-response similarity and dose between pediatric and adult patients.
\newblock \emph{The Journal of Clinical Pharmacology}, 61:\penalty0 165--174, 2021.

\end{thebibliography}


\begin{thebibliography}{99}
\bibitem{Best:2024} Best, N., Ajimi, M., Neuenschwander, B., Saint-Hilary, G. and Wandel, S. (2024), \textit{Beyond
the classical type i error: Bayesian metrics for bayesian designs using informative priors},
Statistics in Biopharmaceutical Research pp. 1--14.

\bibitem{Chen:2000} Chen, M. and Ibrahim, J. G. (2000), \textit{Power prior distributions for regression models}, Statis-
tical Science 15, 46--60.

\bibitem{Chen:2011 } Chen, M., Ibrahim, J. G., Lam, P., Yu, A. and Zhang, Y. (2011), \textit{A simulation-based ap-
proach to bayesian sample size determination for performance under a given model and
for separating models}, Biometrics 67, 1163--1170.

\bibitem{Dette:2024 }Dette, H., Koletzko, L. and Bretz, F. (2024), \textit{Testing for similarity of dose response in multiregional clinical trials}.  \url{https://arxiv.org/pdf/2404.17682}

\bibitem{Dette:2018} Dette, H., Mollenhoff, K., Volgushev, S. and Bretz, F. (2018), \textit{Equivalence of regression
curves}, Journal of the American Statistical Association 113, 711--729.

\bibitem{ICH:2022 } ICH E11(a). (2022), \textit{ICH guideline E11a on pediatric extrapolation (ich) (2022, april 6).}
\url{https://www.ema.europa.eu/en/documents/scientific-guideline/draft-ich-guideline-e11a-pediatric-extrapolation-step-2b_en.pdf}

\bibitem{FDA2008a} FDA (2008a), \textit{FDA-approved drugs}. 
 \url{https://www.accessdata.fda.gov/drugsatfda_docs/label/2008/021976s009lbl.pdf}

\bibitem{FDA2008b} FDA (2008b), Office of clinical pharmacology review.
 \url{https://www.fda.gov/media/75532/download}

\bibitem{Hobbs:2011 } Hobbs, B. P., Carlin, B. P., Mandrekar, S. J. and Sargent, D. J. (2011), \textit{Hierarchical com-
mensurate and power prior models for adaptive incorporation of historical information in
clinical trials}, Biometrics 67, 1047--1056.

\bibitem{Mulb:2022} Mulberg, A. E., Conklin, L. S., Croft, N. M. and for Children Pediatric Extrapolation Work-
ing Group, I.-A. (2022), \textit{Pediatric extrapolation of adult efficacy to children is critical for
efficient and successful drug development}, Gastroenterology 163, 77--83.

\bibitem{Oakley:2023} Oakley, J. and O’Hagan, T. (2023), \textit{SHELF: The Sheffield Elicitation Framework (Version 4)}.
\url{https://shelf.sites.sheffield.ac.uk/}
\bibitem{Park:2008 } Park, T. and Casella, G. (2008), \textit{The Bayesian lasso}, Journal of the American Statistical
Association 103, 681--686.
\bibitem{Psioda:2019 } Psioda, M. and Ibrahim, J. G. (2019), \textit{Bayesian clinical trial design using historical data that
inform the treatment effect}, Biostatistics 20, 400--415.

\bibitem{Schm:2014 } Schmidli, H., Gsteiger, S., Roychoudhury, S., ´OHagan, A., Spiegelhalter, D. and Neuenschwander, B. (2014), \textit{Robust meta-analytic-predictive priors in clinical trials with historical
control information}, Biometrics 70, 1023--1032.

\bibitem{Ye:2023} Ye, J., Zhang, V., Strimenopoulou, F., Zhao, Y., Pan, H., Shabbout, M. and Gamalo, M.
(2023), \textit{Recent use of pediatric extrapolation in pediatric drug development in us}, Journal
of Biopharmaceutical Statistics 33, 681--695.

\bibitem{Yu:2001} Yu, K. and Moyeed, R. A. (2001), \textit{Bayesian quantile regression}, Statistics and Probability
Letters 54, 437--447.
\bibitem{Zhang:2021} Zhang, Q., Travis, J., Rothwell, R., Jay, C. E., Jahidur, R., Zhang, Y., Crentsil, V., Altepeter, T., Lee, J. J., Burckart, G. J., Ganley, C. and Wang, J. (2021), \textit{Applying the
noninferiority paradigm to assess exposure-response similarity and dose between pediatric
and adult patients}, The Journal of Clinical Pharmacology 61, 165--174.
\end{thebibliography}
\begin{comment}
    
\end{comment}

%  If your paper refers to supporting web material, then you MUST
%  include this section!!  See Instructions for Authors at the journal
%  website http://www.biometrics.tibs.org

%\section*{Supporting Information}

%Web Appendix A, referenced in Section~\ref{s:model}, is available with
%this paper at the Biometrics website on Wiley Online
%Library.\vspace*{-8pt}

%\appendix

%  To get the journal style of heading for an appendix, mimic the following.

%\section{}
%\subsection{Title of appendix}

%Put your short appendix here.  Remember, longer appendices are
%possible when presented as Supplementary Web Material.  Please 
%review and follow the journal policy for this material, available
%under Instructions for Authors at \texttt{http://www.biometrics.tibs.org}.

\label{lastpage}

\end{document}